# Numerical and Experimental Studies on the Separation Topology of the MVG Controlled Flow at $M$=2.5

Qin Li [1], Yonghua Yan [2], Ping Lu [3], Adam Pierce [4], Chaoqun Liu [5] and Frank Lu [6]
*University of Texas at Arlington, Arlington, Texas, 76019*

In this paper, the implicitly implemented LES method and fifth order bandwidth-optimized *WENO* scheme are used to make comprehensive studies on the separation topology of the MVG controlled flow at $M$=2.5 and $Re_\theta$=5760. Experiments are also made to verify the prediction of the computation. Analyses are conducted on three categories of the topology: the surface separation, cross-section separation and the three dimensional structure of the vortices. A complete description about the separation topology and a series of new findings are obtained. Among them, a pair of spiral point is first predicted by the computation and verified by the experiment. A corresponding new vortex model with 7 vortex tubes is presented also.

## Nomenclature

| | | |
|---|---|---|
| MVG | = | micro ramp vortex generator |
| $M$ | = | Mach number |
| $Re_\theta$ | = | Reynolds number based on inlet momentum thickness |
| LES | = | large eddy simulation |
| DNS | = | direct numerical simulation |
| 2-D | = | two dimensional |
| 3-D | = | three dimensional |
| SL | = | separation line |
| AL | = | attachment line |
| SP | = | spiral point |
| SDP | = | saddle point |
| HSDP | = | half saddle point |
| NP | = | nodal point |
| HNP | = | half nodal point |
| DVL | = | diverged line |
| CVL | = | converged line |

Subscript
| | | |
|---|---|---|
| $w$ | = | wall |
| $\infty$ | = | free stream |

## I. Introduction

MICRORAMP vortex generator is a kind of passive flow control instrument for boundary layer control. As we know, for shock-boundary layer interaction (SBLI) problems like that in supersonic ramp jets, shock-induced separation can cause total pressure loss, make the flow unsteady and distorting. The worst case can even make the engine unable to start. Because the traditional control technique like bleeding is thought to result in mechanical complexity and be of less efficiency, new alternative like MVG is being intensively studied. In contrary to the conventional vortex generator, MVG has a height approximately 20-40% (more or less) of the boundary layer thickness, which is thought to make the generated streamwise vortex remain in the boundary layer for relatively longer distance. The "down-wash" effect by the streamwise vortices is thought to result in momentum exchange between the fast flow in outer layer and the slow one in the bottom of the boundary layer, and the exchange makes the boundary layer less liable to separate. During such process, a specific phenomenon called as "momentum deficit" will happen[3], i.e., a cylindrical region consisting of low speed flows will be generated after MVG, which is thought to mainly come from the shedding of the boundary layer over MVG and entrained by streamwise vortices[1].

---

[1] Visiting Post Doc., Math Department, 411 S. Nedderman Drive, Arlington, TX 76019-0408, AIAA member.
[2] Ph. D Candidate, Math Department, 411 S. Nedderman Drive, Arlington, TX 76019-0408.
[3] Ph. D Candidate, Math Department, 411 S. Nedderman Drive, Arlington, TX 76019-0408.
[4] Ph. D Candidate, MAE Department, 500 West First Street, Arlington, TX 76019-0018.
[5] Professor, Math Department, 411 S. Nedderman Drive, Arlington, TX 76019-0408, AIAA associate fellow.
[6] Professor, MAE Department, 500 West First Street, Arlington, TX 76019-0018, AIAA associate fellow.

In Ref. 1 and 2, a new phenomenon called as "vortex rings" was first discovered, i.e., a train of vortex rings is generated continuously within the boundary of the momentum deficit. The mechanism for the vortex ring generation was analyzed and found to be that, the existence of the high shear layer caused by the momentum deficit will induce the corresponding Kelvin-Helmholtz instability, which develops into a series of vortex rings[1]. Such process weakens the original streamwise vortices, and makes the dynamics of vortex rings to be at least part of the mechanisms of the flow control. After knowing the computational results and the theoretical analysis[1,2], an experiment was set and corresponding verification has been obtained by the instantaneous image of PIV and acetone vapor by Lu et al[4].

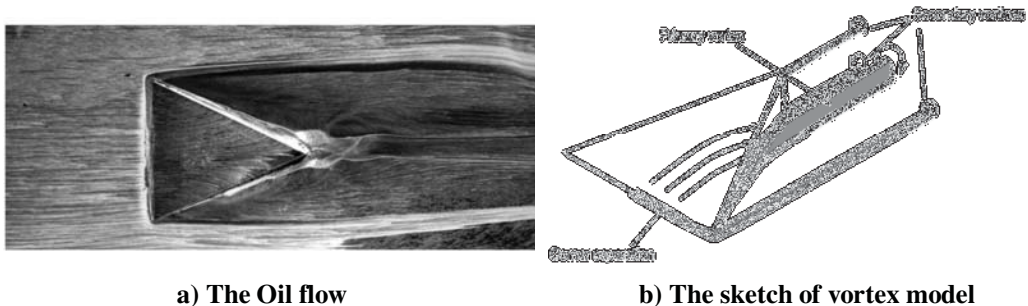

a) The Oil flow            b) The sketch of vortex model

**Figure 1. The oil flow and vortex model from Babinsky[3]**

Although there are vortex rings in the downstream of the MVG flow, the existence of strong streamwise vortices is obvious in the region near MVG. Such flow structures will of course play an important role in the flow evolution and control. So it is necessary to get a deep understanding of the vortex structures and separation topology of the flow near MVG. Due to the lack of comprehensive information, this analysis is far from the state that everything is clear.

Till now, the available clear experimental surface separation pattern is obtained by the experiments of Babinsky[3], in which a series of distinct accumulated oil lines mark the location of separation lines, as shown in Fig. 1(a). According to their experimental results of the flow separation, a vortex model has been derived by Babinsky as follows: the vortices around MVG consist of 4 pairs of vortices, i.e., the horseshoe vortex by the leading edge separation of MVG, the primary vortex by MVG, two secondary vortices under the primary vortex - one lies on the bottom of the plate, the other lies at the side of MVG, as shown in Figure 1(b). The correctness of the model depends on the accuracy of the experiment. In the experiment of Babinsky, the separation line marked by the oil accumulation is clear for the horseshoe vortices, and the main part of the secondary vortices, but the image of oil flow is not distinct near the back end region of MVG. Usually the surface of MVG is poorly visualized by oil flow compared to that on the plate because the size of MVG is relatively small; it is even more difficult to tell the separation structure happened on the side of MVG.

In Ref. 1, we present, for the first time, the possible complex separation topology and the vortex structure caused by MVG. The method is to draw the surface limiting streamlines and deduce the separation pattern. A preliminary vortex model was developed, in which the main part is similar to that of Babinsky, but differences exist in secondary vortex structures. The prominent representative of the difference is the existence of a pair of spiral points predicted by the computation, which indicates a different secondary vortex model. After comparing with experiment by Lu et al and making more deep analyses, more details are discovered and further modification should be added to the model presented in Ref. 1.

According to the current state of researches, the investigation about the separation topology and vortex structures is still immature and some uncertainties exist. This study serves as a try to get deep understanding of vortex structure around MVG.

In the paper, the analyses are made on numerical simulation about the MVG controlled flow at $M=2.5$ and $Re_\theta=5760$. Especially two MVG geometries are studied with the back-edge declining angles of 70° and 45°. In order to obtain more accurate simulations, a high order large eddy simulation method is used to solve the unfiltered form of the Navier-Stokes equations with the 5$^{th}$ order Bandwidth-optimized *WENO* scheme[6], which is generally referred to the so-called implicitly implemented LES[7]. Without explicitly using the subgrid scale (SGS) model as the explicit LES, the implicitly implemented LES uses the intrinsic dissipation of the numerical method to dissipate the turbulent energy accumulated at the unresolved scales with high wave numbers. In order to confirm the computation, an experiment was made for the same problem. Analyses are made about the separation topology and the vortex structure, and discussions are made regarding to the above mentioned purpose.

## II. Numerical Methods

### A. Governing equations

$$\frac{\partial Q}{\partial t}+\frac{\partial E}{\partial x}+\frac{\partial F}{\partial y}+\frac{\partial G}{\partial z}=\frac{\partial E_v}{\partial x}+\frac{\partial F_v}{\partial y}+\frac{\partial G_v}{\partial z}, \qquad (1)$$

where

$$Q=\begin{bmatrix}\rho\\ \rho u\\ \rho v\\ \rho w\\ e\end{bmatrix}\quad E=\begin{bmatrix}\rho u\\ \rho u^2+p\\ \rho uv\\ \rho uw\\ (e+p)u\end{bmatrix}\quad F=\begin{bmatrix}\rho v\\ \rho vu\\ \rho v^2+p\\ \rho vw\\ (e+p)v\end{bmatrix}\quad G=\begin{bmatrix}\rho w\\ \rho wu\\ \rho wv\\ \rho w^2+p\\ (e+p)w\end{bmatrix}$$

$$E_v=\frac{1}{Re}\begin{bmatrix}0\\ \tau_{xx}\\ \tau_{xy}\\ \tau_{xz}\\ u\tau_{xx}+v\tau_{xy}+w\tau_{xz}+q_x\end{bmatrix}\quad F_v=\frac{1}{Re}\begin{bmatrix}0\\ \tau_{yx}\\ \tau_{yy}\\ \tau_{yz}\\ u\tau_{yx}+v\tau_{yy}+w\tau_{yz}+q_y\end{bmatrix}\quad G_v=\frac{1}{Re}\begin{bmatrix}0\\ \tau_{zx}\\ \tau_{zy}\\ \tau_{zz}\\ u\tau_{zx}+v\tau_{zy}+w\tau_{zz}+q_z\end{bmatrix}$$

$$e=\frac{p}{\gamma-1}+\frac{1}{2}\rho(u^2+v^2+w^2)\quad q_x=\frac{\mu}{(\gamma-1)M_\infty^2 \Pr}\frac{\partial T}{\partial x}\quad q_y=\frac{\mu}{(\gamma-1)M_\infty^2 \Pr}\frac{\partial T}{\partial y}$$

$$q_z=\frac{\mu}{(\gamma-1)M_\infty^2 \Pr}\frac{\partial T}{\partial z}\quad p=\frac{1}{\gamma M_\infty^2}\rho T\quad \Pr=0.72$$

$$\tau=\mu\begin{bmatrix}\frac{4}{3}\frac{\partial u}{\partial x}-\frac{2}{3}(\frac{\partial v}{\partial y}+\frac{\partial w}{\partial z}) & \frac{\partial u}{\partial y}+\frac{\partial v}{\partial x} & \frac{\partial u}{\partial z}+\frac{\partial w}{\partial x}\\ \frac{\partial u}{\partial y}+\frac{\partial v}{\partial x} & \frac{4}{3}\frac{\partial v}{\partial y}-\frac{2}{3}(\frac{\partial w}{\partial z}+\frac{\partial u}{\partial x}) & \frac{\partial v}{\partial z}+\frac{\partial w}{\partial y}\\ \frac{\partial u}{\partial z}+\frac{\partial w}{\partial x} & \frac{\partial v}{\partial z}+\frac{\partial w}{\partial y} & \frac{4}{3}\frac{\partial w}{\partial z}-\frac{2}{3}(\frac{\partial u}{\partial x}+\frac{\partial v}{\partial y})\end{bmatrix}$$

The viscous coefficient is given by Sutherland's equation:

$$\mu = T^{\frac{3}{2}} \frac{1+C}{T+C}, \quad C = \frac{110.4}{T_\infty} \qquad (2)$$

The non-dimensional variables are defined as follows:

$$x = \frac{\tilde{x}}{L}, y = \frac{\tilde{y}}{L}, z = \frac{\tilde{z}}{L}, u = \frac{\tilde{u}}{U_\infty}, v = \frac{\tilde{v}}{U_\infty}, w = \frac{\tilde{w}}{U_\infty},$$

$$T = \frac{\tilde{T}}{T_\infty}, \mu = \frac{\tilde{\mu}}{\mu_\infty}, k = \frac{\tilde{k}}{k_\infty}, \rho = \frac{\tilde{\rho}}{\rho_\infty}, p = \frac{\tilde{p}}{\rho_\infty U_\infty^2}, e = \frac{\tilde{e}}{\rho_\infty U_\infty^2}$$

where the variables with '~' are the dimensional counterparts.

Considering the following grid transformation,

$$\begin{cases} \xi = \xi(x, y, z) \\ \eta = \eta(x, y, z) \\ \zeta = \zeta(x, y, z) \end{cases} \qquad (3)$$

the Navier-Stokes equations can be transformed to the system using generalized coordinates:

$$\frac{\partial \hat{Q}}{\partial \tau} + \frac{\partial \hat{E}}{\partial \xi} + \frac{\partial \hat{F}}{\partial \eta} + \frac{\partial \hat{G}}{\partial \zeta} = \frac{\partial \hat{E}_v}{\partial \xi} + \frac{\partial \hat{F}_v}{\partial \eta} + \frac{\partial \hat{G}_v}{\partial \zeta} \qquad (4)$$

where $\hat{Q} = J^{-1}Q$ and

$$\hat{E} = J^{-1}(\xi_x E + \xi_y F + \xi_z G) \quad \hat{F} = J^{-1}(\eta_x E + \eta_y F + \eta_z G) \quad \hat{G} = J^{-1}(\zeta_x E + \zeta_y F + \zeta_z G)$$

$$\hat{E}_v = J^{-1}(\xi_x E_v + \xi_y F_v + \xi_z G_v) \quad \hat{F}_v = J^{-1}(\eta_x E_v + \eta_y F_v + \eta_z G_v) \quad \hat{G}_v = J^{-1}(\zeta_x E_v + \zeta_y F_v + \zeta_z G_v)$$

$J^{-1}, \xi_x$, etc are grid metrics, and $J^{-1} = \det\left(\frac{\partial(x, y, z)}{\partial(\xi, \eta, \zeta)}\right)$, $\xi_x = J(y_\eta z_\zeta - z_\eta y_\zeta)$, etc.

**B. Finite difference schemes and boundary conditions**
*1. The 5th order bandwidth-optimized WENO scheme for the convective terms[6]*
For integrity, the form of the 5th scheme by Weirs and Martin is described as follows. Considering the one dimensional hyperbolic equation:

$$\frac{\partial u}{\partial t} + \frac{\partial f(u)}{\partial x} = 0 \qquad (5)$$

The semi-discretized equation can be expressed as following:

$$\left(\frac{\partial u}{\partial t}\right)_j = -\frac{h_{j+1/2} - h_{j-1/2}}{\Delta x} \quad (6)$$

Considering the positive flux, the four upwind-biased schemes on four candidates can be given as:

$$\begin{cases} h^{+'}_0 = \frac{1}{3}f_{j-2} - \frac{7}{6}f_{j-1} + \frac{11}{6}f_j \\ h^{+'}_1 = -\frac{1}{6}f_{j-1} + \frac{1}{3}f_j + \frac{5}{6}f_{j+1} \\ h^{+'}_2 = \frac{1}{3}f_j + \frac{5}{6}f_{j+1} - \frac{1}{6}f_{j+2} \\ h^{+'}_3 = \frac{11}{6}f_{j+1} - \frac{7}{6}f_{j+2} + \frac{1}{3}f_{j+3} \end{cases} \quad (7)$$

The mark '+' refers to the positive flux after flux splitting. The classic nonlinear weighted schemes by Jiang & Shu[8] can be expressed as:

$$h^+_{j+1/2} = w_0 h^{+'}_0 + w_1 h^{+'}_1 + w_2 h^{+'}_2 + w_3 h^{+'}_3 \quad (8)$$

where $w_i = b_i / \sum_{i=0}^{3} b_i$, $b_i = \alpha_i / (\varepsilon + IS_i)^2$, $(\alpha_0, \alpha_1, \alpha_2, \alpha_3)$ =(0.094647545896, 0.428074212384, 0.408289331408, 0.068988910311), and $\varepsilon$ is a small quantity to prevent the denominator from being zero, which should be small enough in supersonic problems with shocks ($10^{-6} \sim 10^{-10}$). $IS_i$ is the smoothness measurement and has the following form:

$$\begin{cases} IS_0 = \frac{13}{12}(f_{j-2} - 2f_{j-1} + f_j)^2 + \frac{1}{4}(f_{j-2} - 4f_{j-1} + 3f_j)^2 \\ IS_1 = \frac{13}{12}(f_{j-1} - 2f_j + f_{j+1})^2 + \frac{1}{4}(f_{j-1} - f_{j+1})^2 \\ IS_2 = \frac{13}{12}(f_j - 2f_{j+1} + f_{j+2})^2 + \frac{1}{4}(3f_j - 4f_{j+1} + f_{j+2})^2 \\ IS_3 = \frac{13}{12}(f_{j+1} - 2f_{j+2} + f_{j+3})^2 + \frac{1}{4}(-5f_{j+1} + 8f_{j+2} - 3f_{j+3})^2 \end{cases} \quad (9)$$

In order to make the scheme stable, additional modification is made as: $IS_3 = \max_{0 \le k \le 3} IS_k$.

Further improvement for $\omega_k$ made by Martin et al is:

$$\omega_k = \begin{cases} \alpha_k & \text{if } \max(TV_k)/\min(TV_k) < 5 \text{ and } \max(TV_k) < 0.2 \\ \omega_k & \text{otherwise} \end{cases} \quad (10)$$

where $TV_k$ stands for the total variation on each candidate stencil.

The scheme for $h^-_{j+1/2}$ has a symmetric form of $h^+_{j+1/2}$ to the point $x_{j+1/2}$.

*2. The difference scheme for the viscous terms*

Considering the conservative form of the governing equations, the traditional 4$^{th}$ order central scheme is used twice to compute the 2$^{nd}$ order viscous terms.

*3. The time scheme[8]*

The basic methodology for the temporal terms in Navier-Stokes equations adopts the explicit 3$^{rd}$ order TVD-type Runge-Kutta scheme:

$$u^{(1)} = u^n + \Delta t L(u^n)$$
$$u^{(2)} = \frac{3}{4}u^n + \frac{1}{4}u^{(1)} + \frac{1}{4}\Delta t L(u^{(1)})$$
$$u^{n+1} = \frac{1}{3}u^n + \frac{2}{3}u^{(2)} + \frac{2}{3}\Delta t L(u^{(2)})$$
(11)

*4. The inflow conditions and other boundary conditions*

The adiabatic, zero-gradient of pressure and non-slipping conditions are used for the wall as:

$$\partial T/\partial n = 0, \partial p/\partial n = 0, \vec{U} = 0 \quad (12)$$

To enforce the free stream condition, fixed value boundary condition with the free parameters is used on the upper boundary. The validity of the condition is analyzed in Ref. 1. The boundary conditions at the front and back boundary surface in the spanwise direction are treated as the mirror-symmetry condition, which is under the consideration that the problem is about the flow around MVG arrays and only one MVG is simulated.

The outflow boundary conditions are specified as a kind of characteristic-based condition, which can handle the outgoing flow without reflection. The details can be found in Ref. 1.

The inflow conditions are generated using the following steps:

a) A turbulent mean profile is obtained from Ref. 9 for the streamwise velocity (w-velocity) and the distribution is scaled using the local displacement thickness and free stream velocity.

b) The pressure is uniform at inlet and is the same as the free stream value. The temperature profile is obtained using Walz's equation for the adiabatic wall: First the adiabatic wall temperature is determined by $T_w = T_e(1 + r(\gamma-1)/2 \times M_e^2)$, where the subscript 'e' means the edge of the boundary layer and $r$ is the recovery factor with the value 0.9; next the temperature profile is obtained by Walz's equation: $T/T_e = T_w/T_e - r(\gamma-1)/2 \times M_e^2(U/U_e)^2$; thirdly random fluctuations are added on the primitive variables, i.e. $u$, $v$, $w$, $p$, $\rho$. The disturbance has the form: $\varepsilon_{distb} e^{-(y-y_w)^2/\Delta y_{distb}} \times (random - 0.5)/2$, where the subscript 'distb' means the disturbance, "$random$" is the random function with the value between 0~1, $\varepsilon_{distb}$ equals to 0.1 and $\Delta y_{distb}$ equals to $2/3\delta_0$. More details can be seen in Ref. 1-2. A fully developed turbulent inflow is still under development and will be report very soon.

## C. Code validation

Because the 5$^{th}$ order bandwidth-optimized WENO scheme is an already published high order scheme, and applications have been done on DNS problems by the group of Martin, the code validation is only made to check if the algorithm is correctly implemented. The same validation test is made as the one in Ref. 1, and the correctness of the codes is validated.

## III. The surface and cross-section separation topology

As discussed in Ref. 1 in detail, the shedding of the upstream boundary layer over MVG will be entrained by the primary streamwise vortices generated by MVG, which mainly constitute the so-called momentum deficit. The momentum deficit means a region in which the streamwise velocity is far smaller than that of the outside. The spanwise cross-section of the deficit appears in a circular region, and the 3-D view is a cylindrical-like geometry. At the boundary of the deficit, there exists a high shear layer. And the inflection point (1D)/surface (2D) of the velocity shear, produced by the shear layer will result in Kelvin-Helmholtz instability, i.e., the shear layer will lose the stability and vortex rings will be continuously generated. The typical results about the vortex rings, shock wave and other flow structures of the MVG controlled ramp flow can be seen in Fig. 2.

Table 1 gives some information about the grid configuration of the computation. The details about the geometric objects, grid generation, computational domain, etc, are referred in Ref. 1-2 and will not be repeated here.

**Table 1. The geometric parameters for the computation**

| Lx | Ly | Lz | Δx+ | Δy+ | Δz+ |
|---|---|---|---|---|---|
| $3.75\delta_0$ | $5-7.5\delta_0$ | $25.03355\delta_0$ | 26.224 | 1.357-38.376 | 12.788 |

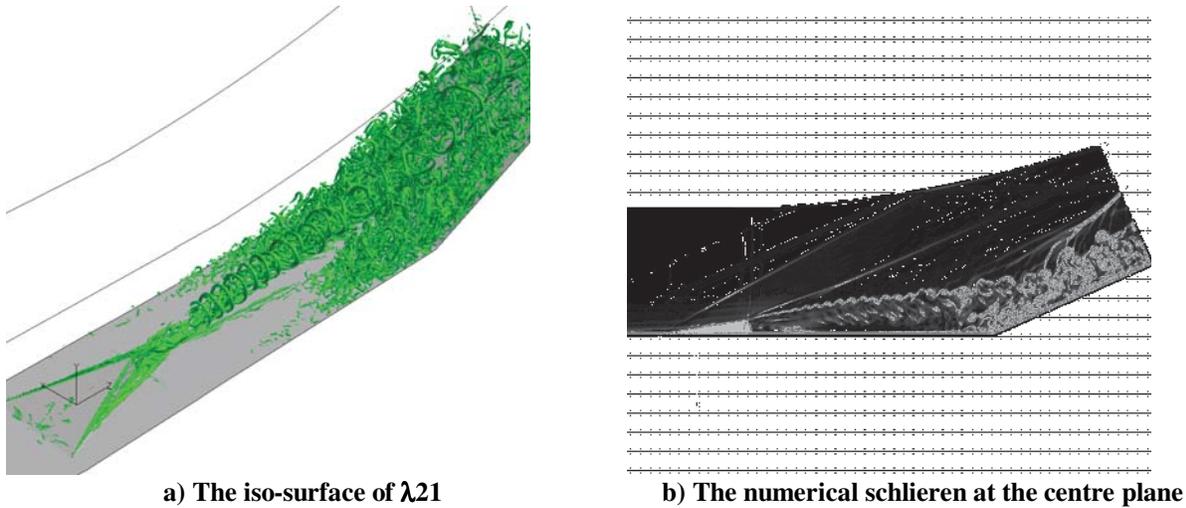

a) The iso-surface of $\lambda_2$1          b) The numerical schlieren at the centre plane
**Figure 2. The typical structures in the flow field**

The purpose of this study is to give a comprehensive investigation of the separation topology and vortex structure in both time-averaged and instantaneous way. Because almost every vortex has its surface footprint before it becomes a fully 3-D one, the surface separation pattern becomes one of the frequent objects in researches. Once the surface pattern is obtained, the 3-D vortex structure can be derived from it. For the experiment in this study, the oil flow is used as a visualization tool and the separation lines are distinguished by the accumulation of the oil; for the computation, the surface limiting streamlines are used to find the separation and attachment lines. And in order to study the 3-D structures of vortices and multiple separations, cross-section separations are also studied by the experiment and computation. The cross-section is generally selected as a plane intersecting with the axis of the vortex. Together with the computation, particle image velocimetry (PIV) and laser beam imaging techniques in the experiment are used to explore the flow structure for verification.

### A. The surface separation topology
*1. The surface separation pattern*

For the MVG controlled flow, the following problems are concerned: a) What is the composition of the separation system, e.g. the number and location of separation lines? b) What is the type of each separation? c) How does it end? d) What are the correspondences between the separation lines and attachment lines? The following analyses are based on the time-averaged results of MVG with the back-edge declining angle of 70°.

a) The framework of the separation system

Considering the approximate symmetry to the center line in statistic sense, only separations on one half-plane of the domain are discussed. Six distinct separation lines and their types can be read form the surface limiting streamlines: *SL1*, the separation line of the leading edge separation of MVG, which corresponds to the horseshoe vortex; *SL2*, the separation of the primary streamwise vortices; *SL3*, one of the *first* secondary separation lines lies on the plate beside MVG; *SL4*, one of the *first* secondary separation lines lies on the side of MVG; *SL5*, another *first* secondary separation lines lies along the back-edge of MVG; *SL8*, the *second* secondary separation line, which will be discussed later. Figures 3-4 provide the locations of the above separation lines.

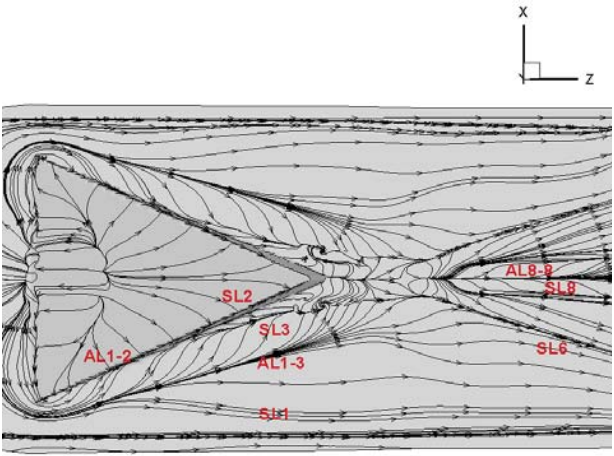

**Figure 3. The separation framework from the top view**

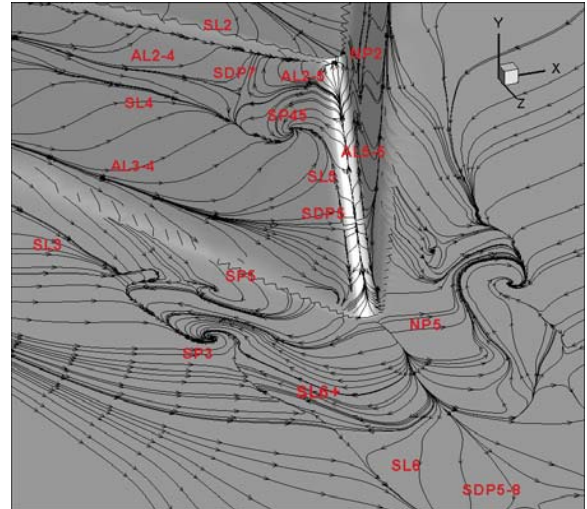

**Figure 4. The separation topology near the foot of MVG**

There might be separations with smaller length scales between the *SL3* and *SL6*, which are labeled as *SL6+* in Fig. 4. Because their length scale is so small, the existence can not be fully validated by the available resolution in current computation. And within the region between *SL2*, *SL4*, the spiral point *SP45* and the nodal point *NP2*, there seems to be another pair of subtle separations, which start from the saddle point *SDP7*. One of separation lines moves down, joins *SL4* and ends in *SP45*, other moves up, joins *SL2* and ends in *NP2*. Similarly, there is a subtle counterpart separation of *SL5* starting at the saddle point *SDP5* but going in downward direction and ending in *SP3*. Because of its small length scale and obscure appearance, it is not labeled in the figure either. The existence of these separations needs further experimental validation.

The horizontal location of the *SP3* is approximately the same as that of the foot of MVG (see Fig. 3). Between the SP3 and MVG, there is another spiral point *SP5* with the horizontal location ahead of that of *SP3* (see Fig. 4). Although it seems to be a questionable small structure, the similar one is found again in the case of 45° MVG, which will be discussed in B.2, and it is the only possible explanation for the "tornado" like structure found by experiment in IV.A.

b) The more details of each separation

*SL1* stands for a closed-type separation, and it extends downstream approximately parallel to the main stream. *SL2* stands for an open-type-started separation, and it ends in a nodal point *NP2*, in which the use of number like '2' follows that of the separation line like *SL2*, as shown in Figs. 4-5. *SL3* also represents an open-type-started separation which starts at the very beginning edge of MVG, and it ends in a spiral point *SP3*. This separation pattern proclaims that, the secondary separation (by *SL3*) terminates and will become a fully 3-D one at *SP3*, which is disconnected and unrelated to other secondary separation like *SL8*. This pattern is different from one by Babinsky[3], who thought *SL3* and *SL8* belong to a same separation line and correspond to a same vortex as well (see Fig.1). *SL4* stands for another open-type-started separation, and it ends in spiral point *SP45*. Because the limiting streamlines below *SL4* move upward (see Fig. 4), it can be expected the experimental oil flow will accumulate above *SL4*, and this is qualitatively consistent to the results of Ford & Babinsky (see Fig.7 in Ref. 5). *SL5* represents a saddle-point-started separation line from *SDP5*, and it ends in the same spiral point *SP45*, as shown in Fig. 4. This separation is a newly discovered one and was not mentioned by the experiment of Babinsky, et al[3, 5]. At last, *SL8* is a node-started (from *NP8*) separation line, and the combination of *NP5*, *SDP5-8* and *NP8* produce a connecting structure between the *first* secondary separation and the *second* secondary separation, as shown in Figs. 3-4. Because *SL3*, *SL4*, and *SL5* are all ended at spiral points, we call them as the *first* secondary separations, and *SL8* are called as the *second* secondary separations which will be discussed later.

Attachment lines are usually located between separation lines. The following obvious attachment lines can be distinguished from the Figs.3-5 as: *AL1-2* separates *SL1* of the horseshoe vortex and *SL2* of the primary streamwise vortex; *AL3-4* separates the two secondary separations lines *SL3* and *SL4*; *AL5-5* separates the secondary separation line *SL5* and its counterpart on the other side of MVG; *AL8-8* separates the secondary separation *SL8* and its symmetric counterpart. The possible *AL2-5* separates the *SL5* and *SL2*.

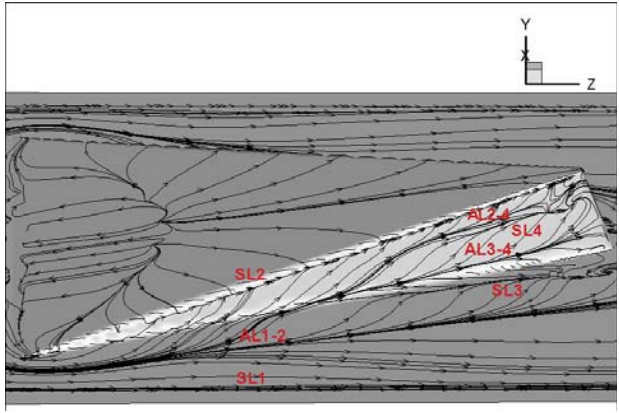

**Figure 5. The separation pattern from the side view**

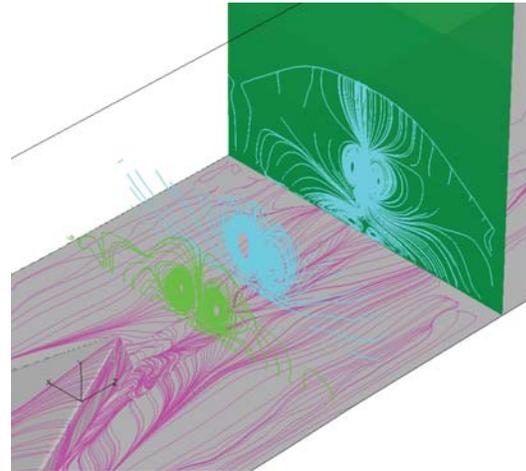

**Figure 6. Instantaneous surface and cross-section streamlines**

Among the separation lines, *SL6* and *SL8* make us very perplexed. From the results of surface oil flow like Fig. 2, Fig. 6 and the above analyses, *SL8* definitely corresponds to a new *second* secondary separation after MVG, but there is no obvious experimental support for *SL6*. In order to further answer this question, the instantaneous cross-section streamlines are drawn at several sections combined with the surface limiting streamlines after MVG in Fig. 6. At the intermediate region between the *first* secondary separation and second one, the first cross-section streamlines show no secondary separation existed; after that, the *second* secondary separation happens with the footprint *SL8*, which is consistent to that drawn in Fig. 3. However, the corresponding secondary vortices are of flat shape and leave the traces in the wall like *SL8*. So strictly speaking, *SL8* should not correspond to other new secondary separations.

*2. The important experimental verification for the separation pattern*

The most eminent difference between the current study and one given by Babinsky about the separation pattern is the existence of the spiral point *SP3*, and the difference directly leads to different secondary vortex models. Known with the computational results, intensive checking was made by experiments[4] at University of Texas at Arlington (UTA). The complete process of the experiment was recorded by video. From the video, it is clearly demonstrated that the spiral points exist just as the computational counterpart qualitatively. And in Fig. 7, one frame of the video was extracted for comparison. The experimental validation provided proofs to the new separation

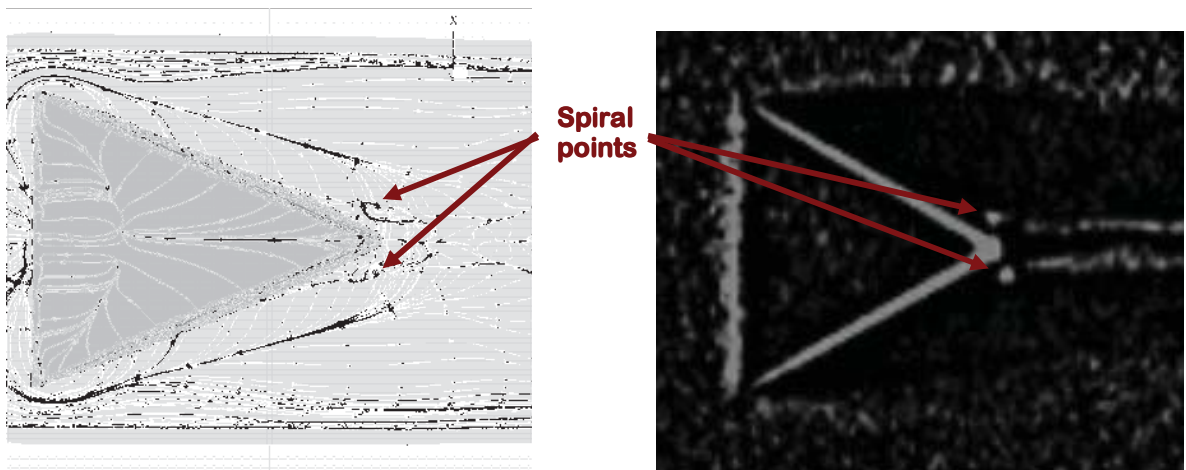

**Figure 7. The spiral points by computation (left) and experiment (right)**

pattern discussed above.

About the existence of separation line *SL5* in Fig. 4, the experimental validation is still expected.

## B. The centre plane and cross-section separation topology

### 1. The centre plane separation topology

The following analyses are based on the time-averaged results.

The central plane is a characteristic plane because it is located in a symmetric position, and the time-averaged velocity on this plane has no spanwise component theoretically. So the flow at this plane is actually a 2-D flow locally. Figure 8 gives the separation topology at the central plane. The dominant structure of the flow is a nodal point *NP*, and the diverging behavior of the point shown by streamlines is apparently caused by the compressibility. And there are two singular points at the boundary. The first is the half saddle point *HSDP1* at the bottom line, which corresponds to the nodal point *NP5* in Fig. 4 and the second is the half saddle point *HSDP2* at the top of MVG, which corresponds to the nodal point *NP2* in Fig. 4. The existence of *NP* indicates the high pressure region nearby as shown in Fig. 8 because the flow speed is slow there. The discrepancy of the position between the singular points and the pressure maximum might be caused by the compressibility.

It can be also inferred from Fig. 8 that some kind of recirculation and dead water region exist just after MVG, and the flow moves downward locally after and below *NP*. After that a diverged line (*DVL*) is gradually generated and divides the flow into two parts: the main part moves upward, which corresponds to the primary streamwise vortices; the other part moves downwards, which corresponds to the secondary vortices. There is an obvious converged line (*CVL*) of the streamlines started from the vertex of MVG. This line is actually the upper boundary of the momentum deficit[1, 2], i.e., a slip line between the fast outer inviscid flow and inner low speed flow. The lower

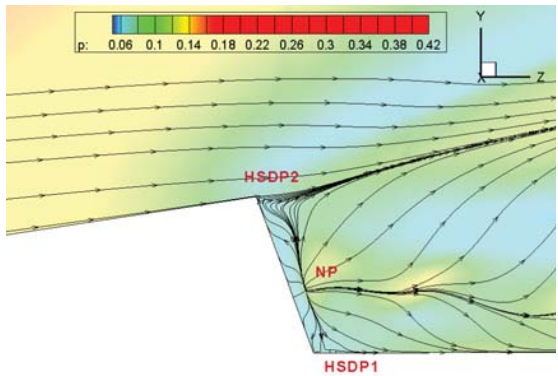 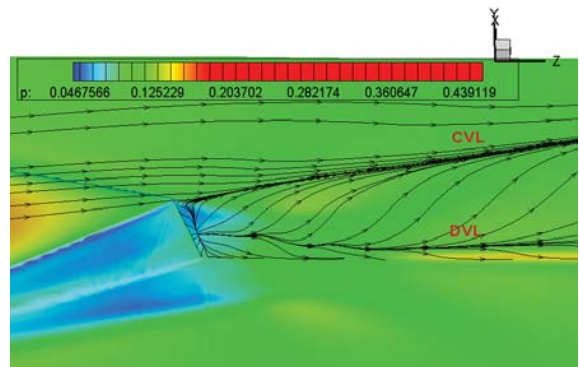

**Figure 8. The separation pattern at the center plane visualized by isobar contour**

**Figure 9. The isobar contour at body surface combined with streamlines at center plane**

boundary of the deficit should be above the diverged line.

Besides MVG with the back edge declining angle 70°, the computation has also been made for MVG with the angle 45° under the same condition. The surface separation pattern is topologically similar to the previous case with an angle of 70°, and differences exist on the size and locations of the structures described above. Especially, the spiral points, the most representative structure, exist for the 45° case as well. By investigating the topology at the center plane, the different structure is found, as shown in Fig. 10. The main separation topology consists of a combination of the saddle point *SDP* and nodal pint *NP*. And there is at least a half nodal point HNP at the center plane which corresponds to the saddle point *SP* in body separation pattern in Fig. 11; a half saddle point *HSDP* in Fig. 10 corresponding to the nodal point *NP5* in Fig. 11; and a half saddle point *HSDP2* in Fig. 10 corresponding to the nodal point *NP2* in Fig. 11. For the surface limiting flow, the flow tends to move downward above *HNP* in Fig.10 or SP in Fig. 11, and move upward below the same point locally. Such behavior will cause the possible oil accumulation above the location of the point in experiment.

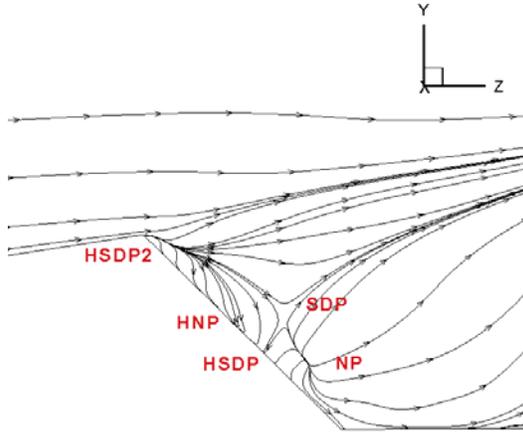 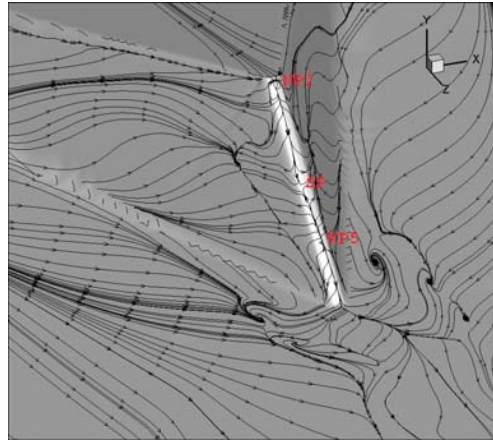

**Figure 10.** The separation pattern at the center plane

**Figure 11.** The separation topology near the foot of MVG

Figure 12 gives the image of the oil flow from the experiment of Ref. 4. It can be observed that the accumulation of the oil above certain line on the side of MVG. Especially, the position of the boundary point of the oil accumulation area on the back-edge of MVG can be determined by the measurement, and the ratio between the length from the top of MVG to the boundary point where the oil vanished and the length of the back-edge is 0.545. In comparison, the computation gives a ratio of 0.57 between the length from the top of MVG to *HNP* in Fig. 10 and the length of the back-edge. The results are pretty close to each other and show a qualitative agreement.

And another consistency worthy to mention between two cases of computation is the existence of inner spiral point at one side of MVG (see Fig. 11), which corresponds to *SP5* in Fig.4. The reason why there is only one additional spiral point is under discussion.

*2. The cross-section separation topology of the second secondary separation*

Based on the analyses in III.A, a cross-section separation topology is presented in Fig. 13 for the region where the two *first* secondary separations occur. In the figure, 'S' represents the location of the start of the separation and 'A' represents that of the attachment. This pattern is similar to that by Babinsky[5] but with more information.

For the structures after MVG, three cross-section separation streamlines are drawn in Fig. 4 and discussion has been made. It is clearly demonstrated that there are two secondary vortices under the primary streamwise vortices. The start point of the secondary separations on the cross-section, which is a half saddle point, is consistent to the separation line on the bottom plate. Because two pairs of vortices are connected by a saddle point, the vortex

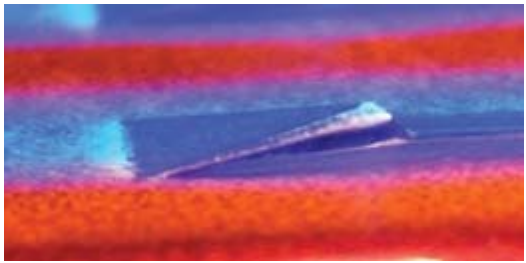 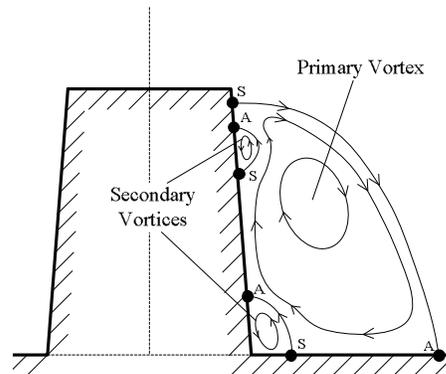

**Figure 12.** The surface oil flow from the experiment[4]

**Figure 13.** The cross-section separation pattern

structure might be unstable and lead to asymmetry.

## IV. The 3-D vortex structure and the revised vortex model

Although detailed information and analyses are given in section III, they are described as topological 2-D structures. However, the corresponding 3-D structures are usually deductions or even conjectures from them. It is necessary to provide direct visualization of the 3-D structures, at least for the less obvious secondary vortices. The topic will be discussed as following, and the analyses are made based on the instantaneous data of the case where the back-edge declining angle of MVG is 70°.

*1. The 3-D structure of the secondary vortices*

Although the structures of the secondary surface separations seem to be complicated in section III, there are three characteristic points which is helpful to clarify the situation, i.e., the spiral points *SP3*, *SP45* and *SP5*. The first two spiral point represents the end of a surface separation and the lift-up of the vortex, e.g., *SP3* corresponds to *SL3* and *SP45* corresponds to *SL4* and *SL5*. In order to further explore the 3-D structure, the following steps are taken: first we choose a computational plane which is away from the body surface and draw the 2-D streamlines on the plane to locate the similar singular point evolved from the original spiral point on the surface; then we put several seeds around the point and investigate the 3-D streamlines originated from the seeds. In Fig. 14 a computational plane near the back-edge of MVG is selected and the topological 2-D streamlines are drawn on that plane. The results show the existence of a similar singular point, which is the continuation of *SP3* on the surface in Fig. 4. Several seeds are put around the point (see Fig. 14), and the corresponding 3-D streamlines are drawn in Fig. 15. Besides the 3-D streamlines, two cross-section streamlines are drawn also to show the primary streamwise vortices. Because the seeds surround the spiral points, the 3-D streamlines started from them approximately reflect the trace of the secondary vortex after the lift-up from the *SP3*. From Fig. 15, the streamlines clearly show that the secondary

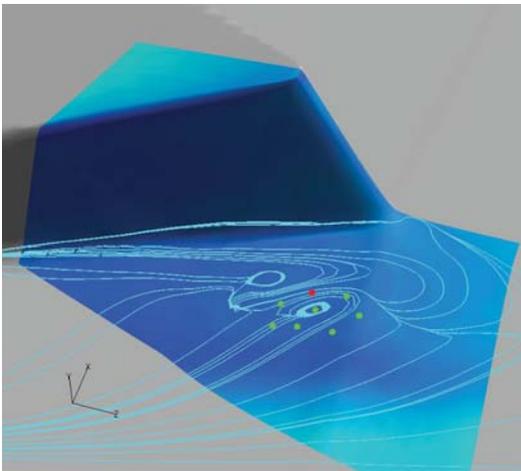 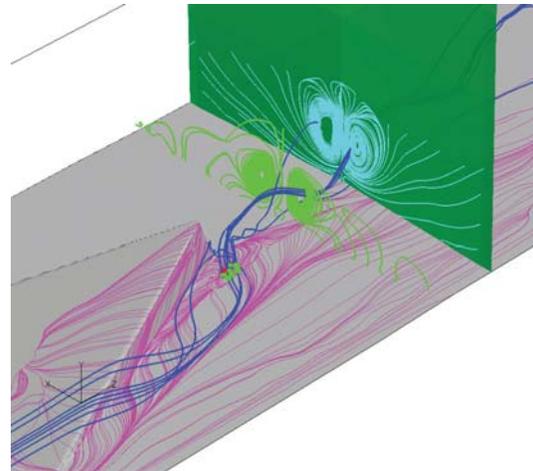

**Figure 14.   The seeds around the continuation of *SP3* on a computational plane away from the body surface**   **Figure 15.   The 3-D streamlines from the seeds and two cross-section streamlines**

vortex is entrained by and rotates around the primary streamwise vortex.

Similar treatment is made for the spiral point *SP45* in Fig. 4. A computational plane away from the body is selected and topological 2-D streamlines are drawn in Fig. 16. A singular point is found as the continuation of *SP45*. Several seeds are placed around the point in order to draw the 3-D streamlines. The trajectories of them indicate that after the surface separation lines *SL4* and *SL5* ending in *SP3* and the vortex lifting up from that point, the secondary vortex is entrained by and rotates around the primary streamwise vortex, as shown in Fig. 17.

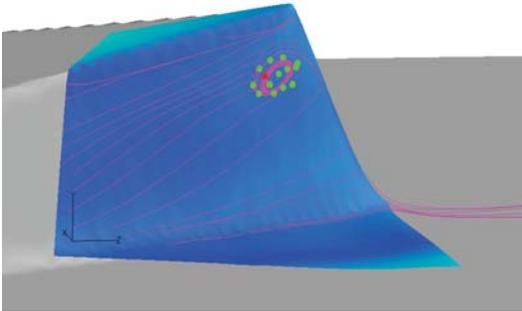

**Figure 16.** The seeds around the continuation of *SP45* on a computational plane away from the wall

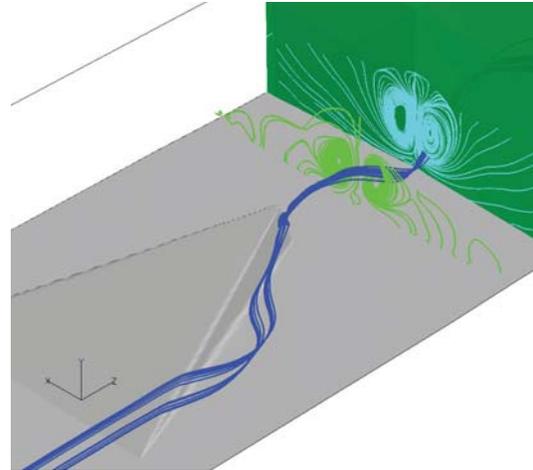

**Figure 17.** The 3-D streamlines from the seeds and two cross-section streamlines

Similarly, investigations are made about the 3-D vortex structure of the second secondary separation represented by *SL8* in Fig. 4. In Fig. 18, after putting several seeds around the vortex core of the secondary vortices, the instantaneous 3-D streamlines originated from the seeds are drawn. It is found that most of 3-D streamlines in Fig. 18 move around the outside of the primary vortices and enter the secondary vortices under the primary ones and without rotation like that in Fig.15 and Fig. 17. This feature demonstrates that from the 3-D view, the vortex corresponding to *SL8* is a newly generated one, and is different from original ones represented by *SL3*, *SL4* and *SL5*.

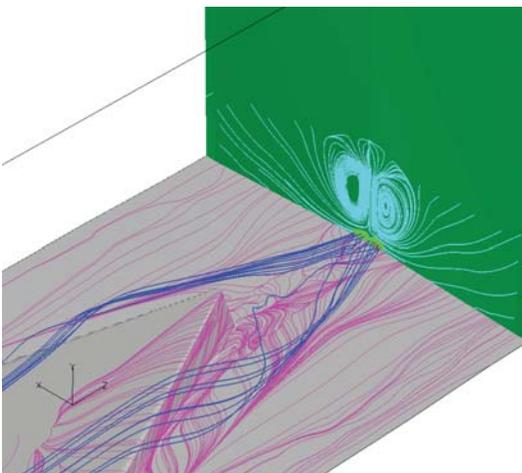

**Figure 18.** The 3-D streamlines from the seeds around the secondary vortex and the cross-section streamlines about the primary vortices

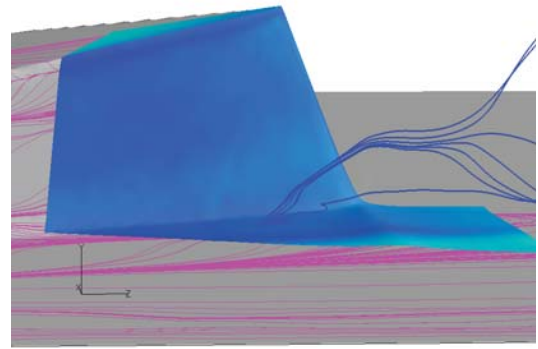

**Figure 19.** The 3-D streamlines originated from the seeds around *SP5*

At last, investigations are made to the 3-D structure emanated from the spiral points *SP5* using similar technique. Figure 19 depicts the 3-D streamlines originated from the seeds around the inner center structure (the continuation of *SP5*) at the computational plane which is located at 8 layers away from the wall (see Fig. 14). From the figure, we can see that the streamlines move up aslant at an angle of around 45°. Recently, a movie was taken from the side view of MVG in an experiment at UTA. A so-called "tornado"-like vortex filament was discovered by the movie. Such structure is found to be very stable and robust during the period of the wind tunnel test. A snapshot was taken from the video and shown in Fig. 20, in which the vortex filament is the bright line at an angle of around 50°. It is necessary to point out that, the filament is not the one originated from the main spiral point *SP3*, because the streamwise coordinates of *SP3* is almost the same as that of the foot of MVG (see Fig.7), and the lift up of vortex from *SP3* has a different position from that shown in Fig.20. So the only possibility is that the filament comes from the vortex lift-up of the spiral point *SP5*. Although there is still some discrepancies on the location and the angle of the vortex between the computation and experiment, a qualitative agreement should be considered as obtained, which can be used to show the existence of *SP5*.

   *2. The revised vortex model*

   Based on the above analyses, a revised vortex model is presented as shown in Fig. 21, in which only the vortices with considerable size are considered. The framework of the vortices is composed of the horseshoe vortex, the primary streamwise vortex, the *first* secondary vortices and the *second* secondary vortices. The *first* secondary vortices consist of one vortex lying on the plate with the separation line *SL3* in Fig. 4-5, and two vortices lying on the side of MVG with the respective separation lines *SL4* and *SL5* in Fig. 4. The one corresponding to *SL3* becomes a 3-D vortex by ending at the spiral point *SP3*, and the rest two vortices become a 3-D vortex by ending at the spiral point *SP45*. The *second* secondary vortex is a newly generated vortex underneath the primary streamwise vortex with the separation line *SL8*. Because of the limited space, the sketch for vortex from *SP5* is not drawn in the figure.

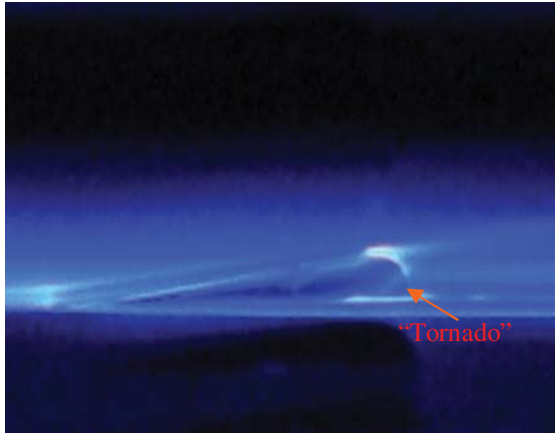 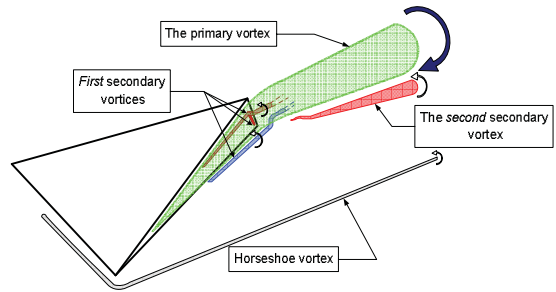

**Figure 20.   The vortex filament discovered at the side of MVG by the experiment**

**Figure 21.   The revised vortex model shown in half of the domain**

The total number of the vortices is seven when counting in the half domain.

## V.   Discussions about the leading edge separation of MVG

   Because the experiments are of high importance to verify the predictions of the computation and analyses, it is first needed to guarantee the correctness of the experiment. By taking the oil flow movie of MVG flow in one experiment at UTA, it is found that different results are obtained at the different running stages of the wind tunnel. Figure 22 shows the snapshot at the start up stage of the wind tunnel. At this stage, the oil flow appears in filaments with high resolution, and the oil flows quite fast when watched from the movie. What is more, there is no visible separation ahead of MVG (see Fig. 22).  When the wind tunnel runs stable and the spiral points *SP3s* appear, there is obvious oil accumulation ahead of MVG (see Fig. 23), which indicates the leading edge separation. There are even some hints of attachment line on MVG surface. The oil flows lose the filament characteristic and have the particle-like appearance; the speed of oil flows becomes rather slow as well. After the stable running stage, the wind tunnel is going to be shut down. Some features of the Fig. 23 are lost at this stage.

   According to the movie made by our experiment, there is definite leading edge separation with certain size. In order to experimentally determine the size of leading edge separation as exact as possible, it is necessary to use the dynamic video recording technique and further enhance the resolution of video frames.

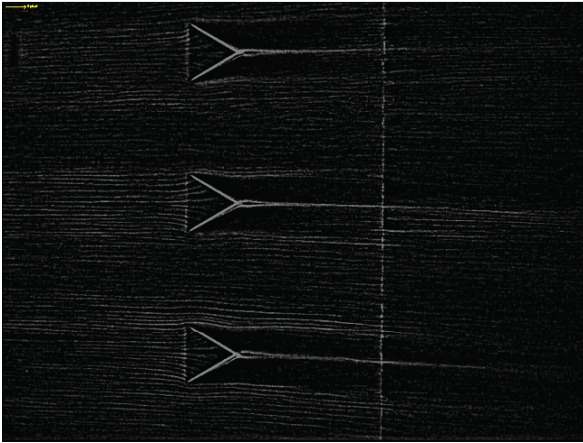 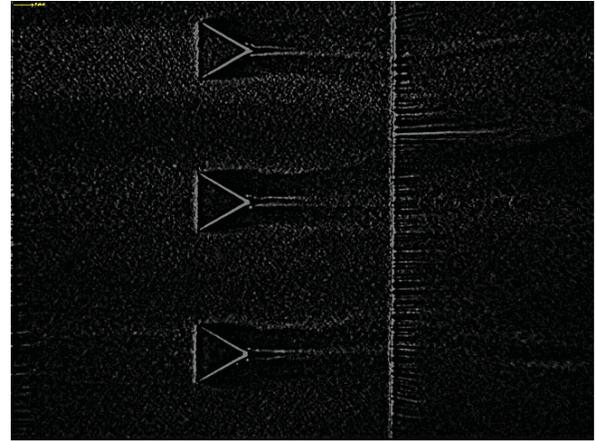

**Figure 22.** The oil flow pattern during the start up stage of the wind tunnel

**Figure 23.** The oil flow pattern during the stable running stage of the wind tunnel

## VI. Conclusion

In this paper, the detailed analyses are made on the separation topology of the MVG controlled flow at M=2.5 and $Re_\theta$=5760. The analyses are made based on the simulation using implicitly implemented LES method, in which the 5$^{th}$ order *WENO-type* scheme developed by Weirs & P. Martin is used. Experimental verifications are made using the results from Ref. 4. The following conclusions can be made:

1. The complete knowledge about the surface separation pattern is abstained, including the location, the type, and the start and end information of each separation. Especially, a pair of spiral points, i.e. *SP3s*, is first discovered by the computation, and verified by the experiment later. The importance of the finding is that it indicates a new secondary separation model. The validation about the other structures is still open for the validation by experiments, like the spiral point *SP45*, *SP5* and the secondary separation *SL5*, etc.
2. The separation pattern about centre plane and downstream cross-section is revealed by analyses. The topology of the centre plane and cross-section separation is consistent to that obtained from the surface separation.
3. The 3-D structures of the vortices are studied and used to verify the vortex structure derived from surface separations. The secondary vortex corresponding to the separation *SL3* is independent from that corresponding to *SL4* and *SL5*, and they are all entrained by and rotate around the primary streamwise vortex. The *second* secondary vortex corresponding to the separation line *SL8* is unrelated to the *first* secondary vortices (with respect to *SL3, SL4* and *SL5*) and is located under the primary streamwise vortex.
4. A revised vortex model is presented in the paper different from that of Babinsky[3], which has 7 vortices in half domain, i.e., one horseshoe vortex, one primary streamwise vortex, four *first* secondary vortices and one *second* secondary vortices.

### Acknowledgements

This work is supported by AFOSR grant FA9550-08-1-0201 supervised by Dr. John Schmisseur. The authors are grateful to Texas Advantage Computing Center (TACC) for providing computation hours.